\documentstyle[pra,aps,twocolumn,fleqn,epsfig]{revtex}

\begin{document}
\draft

\twocolumn[\hsize\textwidth\columnwidth\hsize\csname@twocolumnfalse\endcsname

\title{Effects of anisotropy on thermal entanglement }
\author{Xiaoguang Wang}
\address{Institute of Physics and Astronomy, Aarhus University, 
         DK-8000, Aarhus C, Denmark}
\date{\today}
\maketitle

\begin{abstract}
We study the thermal entanglement in the two-qubit anisotropic XXZ model and the 
Heisenberg model with Dzyaloshinski-Moriya (DM) interactions. The DM interaction is another
kind of anisotropic antisymmetric exchange interaction. The effects of these two kinds of 
anisotropies on the thermal entanglement are studied in detail for  both the antiferromagnetic and 
ferromagnetic cases.
\end{abstract}

]

Recently the concept of thermal entanglement was introduced and studied
within one-dimensional isotropic Heisenberg model\cite{Arnesen}. The state of the
system described by the Hamiltonian $H$ at thermal equilibrium is $\rho
(T)=\exp \left( -\frac H{kT}\right) /Z,$ where $Z=$Tr$\left[ \exp \left( -%
\frac H{kT}\right) \right] $ is the partition function and $k$ is the
Boltzmann$^{\prime }$s constant. As $\rho (T)$ represents a thermal state, the
entanglement in the state is called the thermal entanglement\cite{Arnesen}.

For two-qubit isotropic Heisenberg model there exists thermal entanglement
for the antiferromagnetic case and no thermal entanglement for the
ferromagnetic case\cite{Arnesen}. While for the $XY$ model 
the thermal entanglement appears for
both the antiferromagnetic and ferromagnetic cases\cite{Wang}. It is known
that the isotropic Heisenberg model and the $XY$ model are special cases of
the anisotropic Heisenberg model (see Eq.(\ref{eq:hhh})). So it is worth to
study the thermal entanglement in the anisotropic models and see the role 
of anisotropic parameters. In this paper we
consider two types of anisotropy and study the effects of
them on the thermal entanglement. Both the antiferromagnetic and
ferromagnetic cases are considered.

First we briefly review a measure of entanglement, the concurrence\cite{Wootters1}.
Let $\rho _{12}$ be the density matrix of a pair of qubits $1$ and $2.$ The
density matrix can be either pure or mixed. The concurrence corresponding to
the density matrix is defined as

\begin{equation}
C_{12}=\max \left\{ \lambda _1-\lambda _2-\lambda _3-\lambda _4,0\right\},
\label{eq:c1}
\end{equation}
where the quantities $\lambda _1\ge \lambda _2\ge \lambda _3\ge $ $\lambda
_4 $ are the square roots of the eigenvalues of the operator 
\begin{equation}
\varrho_{12} =\rho _{12}(\sigma _{1y}\otimes \sigma _{2y})\rho _{12}^{*}(\sigma
_{1y}\otimes \sigma _{2y}).  \label{eq:c2}
\end{equation}
The opeators $\sigma _{jy} (j=1,2)$ are the usual Pauli operators for the qubit $j$.
The concurrence $C_{12}=0$ corresponds to an unentangled state and $C_{12}=1$
corresponds to a maximally entangled state.

We consider the two-qubit anisotropic XXZ Heisenberg model\cite{Bethe,XXZ}

\begin{eqnarray}
H &=&\frac J2\left( \sigma _{1x}\sigma _{2x}+\sigma _{1y}\sigma _{2y}+\Delta
\sigma _{1z}\sigma _{2z}\right)   \nonumber \\
&=&J\left( \sigma _{1+}\sigma _{2-}+\sigma _{1-}\sigma _{2+}\right) +\frac{%
J\Delta }2\sigma _{1z}\sigma _{2z},  \label{eq:hhh}
\end{eqnarray}
where the coupling constants $J>0$ corresponds to the antiferromagnetic case and 
$J<0$ the ferromagnetic case. The operators $\sigma _{j\pm }=\frac 12\left(
\sigma _{jx}\pm i\sigma _{jy}\right) (j=1,2).$ The XXZ model was initiated by
Bethe for the case $\Delta=\pm 1$ in 1931\cite{Bethe} and has been studied for $\Delta\ne\pm1$
since 1959\cite{XXZ}.

The eigenvalues and
eigenvectors of $H$ are easily obtained as

\begin{eqnarray}
H|00\rangle &=&\frac{J\Delta }2|00\rangle ,H|11\rangle =\frac{J\Delta }2%
|11\rangle ,  \nonumber \\
H|\Psi ^{\pm }\rangle &=&\left( -\frac{J\Delta }2\pm J\right) |\Psi ^{\pm
}\rangle ,  \label{eq:evals}
\end{eqnarray}
where $|\Psi ^{\pm }\rangle =\frac 1{\sqrt{2}}(|01\rangle \pm |10\rangle )$
are maximally entangled states and $|0\rangle$ ($|1\rangle$) denotes the ground
(excited) state of a two-level particle.

In the standard basis, $\left\{ |00\rangle ,|01\rangle ,|10\rangle
,|11\rangle \right\} ,$ the density matrix $\rho (T)$ is written as ($k=1$)

\begin{equation}
\rho (T)=\frac 1{2(e^{\frac{J\Delta }{2T}}\cosh \frac JT+e^{-\frac{J\Delta }{%
2T}})}
\end{equation}

\begin{center}
$\times \left( 
\begin{array}{llll}
e^{-\frac{J\Delta }{2T}} & 0 & 0 & 0 \\ 
0 & e^{\frac{J\Delta }{2T}}\cosh \frac JT & -e^{\frac{J\Delta }{2T}}\sinh 
\frac JT & 0 \\ 
0 & -e^{\frac{J\Delta }{2T}}\sinh \frac JT & e^{\frac{J\Delta }{2T}}\cosh 
\frac JT & 0 \\ 
0 & 0 & 0 & e^{-\frac{J\Delta }{2T}}
\end{array}
\right). $
\end{center}

The square roots of the four eigenvalues of the density matrix $\varrho _{12}$
are

\begin{eqnarray}
\lambda _1 &=&\lambda _2=\frac{e^{-\frac{J\Delta }T}}{2(\cosh \frac JT+e^{-%
\frac{J\Delta }T})}\text{ },  \nonumber \\
\lambda _3 &=&\frac{e^{\frac JT}}{2(\cosh \frac JT+e^{-\frac{J\Delta }T})}, 
\nonumber \\
\lambda _4 &=&\frac{e^{-\frac JT}}{2(\cosh \frac JT+e^{-\frac{J\Delta }T})}.
\end{eqnarray}

Which is the largest eigenvalue depends on the value of anisotropic parameter $%
\Delta $ and sign of $J.$ For antiferromagnetic case ($J>0$) the largest
eigenvalue is $\lambda _1$ when $\Delta \leq -1$ and $\lambda _3$ when $%
\Delta >-1.$ Therefore the concurrences are given by

\begin{eqnarray}
C_{AFM}(\Delta ) &=&0\text{ for }\Delta \leq -1,  \nonumber \\
C_{AFM}(\Delta ) &=&\max \left( \frac{\sinh (\frac JT)-e^{-\frac{J\Delta }T}%
}{\cosh \frac JT+e^{-\frac{J\Delta }T}},0\right) \nonumber\\
\text{ for }\Delta >-1.
\label{eq:cafm}
\end{eqnarray}
When $\Delta =1,$ the anisotropic model becomes the isotropic model, and
Eq.(\ref{eq:cafm}) reduces to

\begin{equation}
C_{AFM}(1)=\max \left( \frac{e^{\frac{2J}T}-3}{e^{\frac{2J}T}+3},0\right) 
\end{equation}
which is obtained in Ref.\cite{Arnesen}. From the above equation we know
that when the temperature is larger than the critical temperature $T_C=\frac{%
2J}{\ln 3}$ the thermal entanglement disappears. For the anisotropic model
the critical temperature $T_C$ is determined by the nonlinear equation 
\begin{equation}
\sinh (\frac JT)=e^{-\frac{J\Delta }T}.  \label{eq:afmeq}
\end{equation}

For ferromagnetic case ($J<0$) the largest eigenvalue is $\lambda _4$ when $%
\Delta <1$ and $\lambda _1$ when $\Delta \geq 1.$ Therefore the concurrences
are

\begin{eqnarray}
C_{FM}(\Delta ) &=&0\text{ for }\Delta \geq 1,  \nonumber \\
C_{FM}(\Delta ) &=&\max \left( \frac{\sinh (\frac{|J|}T)-e^{\frac{|J|\Delta }%
T}}{\cosh \frac{|J|}T+e^{\frac{|J|\Delta }T}},0\text{ }\right)\nonumber\\
\text{for }\Delta <1.  \label{eq:cfm}
\end{eqnarray}
From the above equation we see that no thermal entanglement for the
ferromagnetic isotropic Heisenberg model ($\Delta=1$). The critical temperature is given
by the equation

\begin{equation}
\sinh (\frac{|J|}T)=e^{\frac{|J|\Delta }T}.  \label{eq:fmeq}
\end{equation}

From Eqs.(\ref{eq:cafm}) and (\ref{eq:cfm}) it is find that the thermal
entanglement are same when $\Delta =0.$ That is to say, the entanglement
exists in the antiferromagnetic and ferromagnetic models at the same time.
The Heisenberg Hamiltonian with $\Delta =0$ is just the quantum $XY$ model.
The thermal entanglement this model is discussed in a recent paper%
\cite{Wang}. From Eqs.(\ref{eq:cafm}) and (\ref{eq:cfm}) we also see that the concurrences
satisfy $C_{AFM}(\Delta )=C_{FM}(-\Delta ).$

We numerically solved Eqs.(\ref{eq:afmeq}) and (\ref{eq:fmeq}) and the
results are shown in Fig.1. For the antiferromagnetic case we observe that
the critical temperature $T_C$ increases as the anisotropic parameter $%
\Delta $ increases. Oppositely $T_C$ decreases as $\Delta $ increases for
the ferromagnetic case. Of course the critical temperatures are same when $%
\Delta =0,$ which corresponding to the $XY$ model. 

Fig.2(a) gives a plot of the concurrence
as a function of temperature for the antiferromagnetic case. It
shows that the concurrences are 1 for  different anisotropic parameters when 
$T=0.$ In these cases the ground state is $|\Psi ^{-}\rangle ,$ which is the
maximally entangled state and the corresponding concurrences are 1. As the
temperature increases, the concurrence decreases due to the mixing of other
states with the maximally entangled state. Again we see that $T_C$ increases
as $\Delta $ increases.
\begin{figure}
\epsfig{width=8cm,file=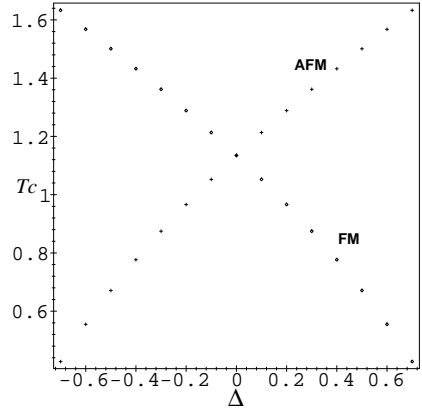}
\caption[]{The critical temperature as a function of anisotropic parameter $%
\Delta $ for both the antiferromagnetic and ferromagnetic cases. The parameter $J=1$.
}
\end{figure}

Another kind of anisotropy is the DM anisotropic antisymmetric interaction
which arises from spin-orbit coupling\cite{D,M}. Now we consider the
Heisenberg model with DM interaction

\begin{eqnarray}
H_{DM} &=&\frac J2[\left( \sigma _{1x}\sigma _{2x}+\sigma _{1y}\sigma
_{2y}+\Delta \sigma _{1z}\sigma _{2z}\right)   \nonumber \\
&&+\vec{D}\cdot (\vec{\sigma}_1\times \vec{\sigma}_2)],
\end{eqnarray}
where $\vec{D}$ is the DM vector coupling. To see the effect of the
anisotropic parameter $\vec{D}$ we choose $\vec{D}=D\vec{z}$ and $\Delta =0$%
. Then the Hamiltonian $H_{DM}$ becomes

\begin{eqnarray}
H_{DM} &=&\frac J2[\sigma _{1x}\sigma _{2x}+\sigma _{1y}\sigma
_{2y}+D(\sigma _{1x}\sigma _{2y}-\sigma _{1y}\sigma _{2x})]  \nonumber \\
&=&J[(1+iD)\sigma _{1+}\sigma _{2-}+(1-iD)\sigma _{1-}\sigma _{2+}].
\end{eqnarray}
The eigenvalues and eigenvectors of $H_{DM}$ are given by

\begin{eqnarray}
H_{DM}|00\rangle &=&0,H_{DM}|11\rangle =0,  \nonumber \\
H_{DM}|\pm \rangle &=&\pm J\sqrt{1+D^2}|\pm \rangle ,
\end{eqnarray}
where $|\pm \rangle =\frac 1{\sqrt{2}}\left( |01\rangle \pm e^{i\theta
}|10\rangle \right) $ and $\theta =\arctan D.$

In the standard basis, the density matrix $\rho (T)$ is given by

\begin{equation}
\rho (T)=\frac 1{2(\cosh \frac{J\sqrt{1+D^2}}T+1)}
\end{equation}

\begin{center}
$\times \left( 
\begin{array}{llll}
1 & 0 & 0 & 0 \\ 
0 & \cosh \frac{J\sqrt{1+D^2}}T & -e^{-i\theta }\sinh \frac{J\sqrt{1+D^2}}T
& 0 \\ 
0 & -e^{i\theta }\sinh \frac{J\sqrt{1+D^2}}T & \cosh \frac{J\sqrt{1+D^2}}T & 
0 \\ 
0 & 0 & 0 & 1
\end{array}
\right). $
\end{center}

The square roots of the four eigenvalues of the density matrix $\varrho _{12}$
are

\begin{eqnarray}
\lambda _1 &=&\lambda _2=\frac 1{2(\cosh \frac{J\sqrt{1+D^2}}T+1)}\text{ ,} 
\nonumber \\
\lambda _3 &=&\frac{e^{\frac{J\sqrt{1+D^2}}T}}{2(\cosh \frac{J\sqrt{1+D^2}}T%
+1)}, \nonumber \\
\lambda _4 &=&\frac{e^{-\frac{J\sqrt{1+D^2}}T}}{2(\cosh \frac{J\sqrt{1+D^2}}T%
+1)}.
\end{eqnarray}
We see that the four eigenvalues are independent on the angle $\theta .$
From the eigenvalues we observe that for both antiferromagnetic and
ferromagnetic cases the concurrences are given by

\begin{equation}
C=\max \left( \frac{\sinh \frac{|J|\sqrt{1+D^2}}T-1}{\cosh \frac{|J|\sqrt{%
1+D^2}}T+1},0\right).
\end{equation}
We see that the entanglement does not depend on the
sign of the anisotropic parameter $D$. 

The critical temperature is given by 
\begin{equation}
T_C=\frac{|J|\sqrt{1+D^2}}{\arcsin \text{h}(1)}\approx 1.1346\sqrt{1+D^2}|J|.
\end{equation}
Obviously the critical temperature increases with the increase of the
absolute value of the anisotropic parameter $D,$ which can also be seen in Fig.2(b). The
ground state of the system is $|+\rangle $ or $|-\rangle $ no matter what
the anisotropic parameter $D$ is. Fig.2(b) shows that the concurrences are 1
when $T=0,$ which is due to the maximally entangled ground state $|\pm
\rangle .$ When the temperature is larger than the critical temperature 
the thermal entanglement disappears. 
\begin{figure}
\epsfig{width=8cm,file=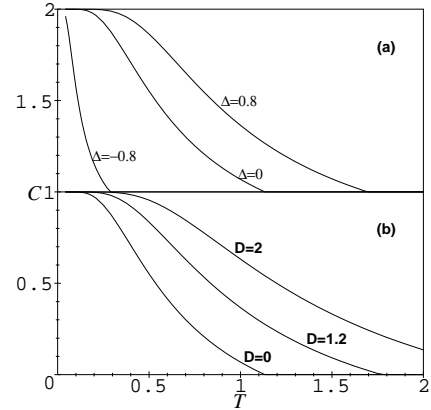}
\caption[]{The concurrences as a function of temperature. (a) For anisotropic
antiferromagnetic Heisenberg model; (b)For Heisenberg model with DM
interaction. The parameter $|J|=1$.}
\end{figure}

In conclusion we have studied the effect of two kinds of anisotropy on the
thermal entanglement in the anisotropic XXZ model and the Heisenberg
model with DM interaction. For the XXZ model it is shown that the thermal 
entanglement exist or not depends on both the anisotropic parameters and the sign of exchange
constants $J.$  The thermal entanglement are same for the antiferromagnetic
and ferromagnetic Heisenberg model with DM interaction. While in the
XXZ model the thermal entanglements are different for the
antiferromagnetic and ferromagnetic cases. In this paper we restrict ourselves to the two-qubit case.
It is a good challenge to study thermal entanglement in the multi-qubit anisotropic models.

\acknowledgments

The author thanks Klaus M\o lmer and Anders S\o rensen for many valuable
discussions with them. This work is supported by the Information Society 
Technologies Programme IST-1999-11053, EQUIP, action line 6-2-1.


\begin{references}
\bibitem{Arnesen}  M. C. Arnesen, S. Bose, and V. Vedral, quant-ph/0009060;
M. A. Nielsen, quant-ph/0011036.

\bibitem{Wang}  X. Wang, quant-ph/0101013.

\bibitem{Wootters1}  S. Hill and W. K. Wootters, Phys. Rev. Lett. {\bf 78},
5022 (1997); W. K. Wootters, Phys. Rev. Lett. {\bf 80}, 2245 (1998); V. Coffman, J.
Kundu, and W. K. Wootters, \pra {\bf 61}, 052306 (2000).

\bibitem{Bethe}H. A. Bethe, Z. Physik {\bf 71}, 205 (1931).

\bibitem{XXZ}R. Orbach, Phys. Rev. {\bf 112}, 309 (1958); 
             L. R. Walker, Phys. Rev. {\bf 116}, 1089 (1959);
             J. des Cloizeaux and M. Gaudin, J. Math. Phys. {\bf 7}, 1384 (1966);
             C. N. Yang and C. P. Yang, Phys. Rev. {\bf 150}, 321 (1966) 

\bibitem{D}  I. Dzyaloshinskii, J. Phys. Chem. Solids {\bf 4}, 241 (1958).

\bibitem{M}  T. Moriya, Phys. Rev. Lett. {\bf 4}, 228 (1960); Phys. Rev. 
{\bf 117}, 635 (1960); {\bf 120}, 91 (1960);
\end{references}
\end{document}